\documentclass[fleqn,12pt,twoside]{article}
\usepackage[headings]{espcrc1}

\readRCS
$Id: espcrc1.tex,v 1.2 2004/02/24 11:22:11 spepping Exp $
\usepackage{graphicx}
\usepackage[figuresright]{rotating}

\newcommand{\AmS}{{\protect\the\textfont2
  A\kern-.1667em\lower.5ex\hbox{M}\kern-.125emS}}

\title{Towards the critical behavior for the light nuclei by NIMROD detector}

\author{Y. G. Ma\address[SINAP]{Shanghai Institute of Applied Physics, Chinese
Academy of Sciences,  China}\address[TAMU]{Cyclotron Institute,
Texas A\&M University, College Station, Texas, USA}, J. B.
Natowitz\addressmark[TAMU], R. Wada\addressmark[TAMU], K.
Hagel\addressmark[TAMU], J. Wang\addressmark[TAMU], T. Keutgen
\address[Belgium]{UCL, Louvain-la-Neuve, Belgium}, Z. Majka
\address[Poland]{Jagiellonian University, Krakow, Poland}, M.
Murray\address{University of Kansas, Lawrence KS 66045, USA}, L.
Qin\addressmark[TAMU], P. Smith\addressmark[TAMU],  R.
Alfaro\address[Mexico]{Instituto de Fisica, UNAM, Mexico City,
Mexico}, J. Cibor\address{Institute of Nuclear Physics, Krakow,
Poland}, M. Cinausero\address[Legnaro]{INFN, Laboratori Nazionali
di Legnaro, I-35020 Legnaro, Italy  }, Y. El Masri
\addressmark[Belgium],
D. Fabris\address[Padova]{INFN and Dipartimento di Fisica
dell'Universit\'a di Padova, I-35131 Padova, Italy  }, E. Fioretto
\addressmark[Legnaro],
 A. Keksis\addressmark[TMAU], M. Lunardon\addressmark[Padova],
 A. Makeev\addressmark[TAMU], N.
Marie\address{LPC, IN2P3-CNRS, ISMRA et Universit\'e, F-14050 Caen
Cedex, France},
 E. Martin\addressmark[TAMU],
A. Martinez-Davalos\addressmark[Mexico], A.
Menchaca-Rocha\addressmark[Mexico], G. Nebbia
\addressmark[Padova], G. Prete\addressmark[Legnaro], V. Rizzi
\addressmark[Padova], A. Ruangma\addressmark[TAMU], D. V.
Shetty\addressmark[TAMU], G. Souliotis\addressmark[TAMU], P.
Staszel\addressmark[Poland], M. Veselsky\addressmark[TAMU], G.
Viesti\addressmark[Padova], E. M. Winchester\addressmark[TAMU], S.
J. Yennello\addressmark[TAMU]
         }

\runtitle{}

\runauthor{}

\begin{document}

\maketitle

\begin{abstract}
The critical behavior for the light nuclei with A$\sim 36$  has
been investigated experimentally by the NIMROD multi-detectors.
The  wide variety of observables indicate the critical point has
been reached in the disassembly of hot nuclei  at an excitation
energy of 5.6$\pm$0.5 MeV/u.
\end{abstract}

\section{INTRODUCTION}
Most efforts to determine the critical point for the expected
liquid gas-phase transition in finite nucleonic matter have
focused on examinations of the temperature and excitation energy
region where maximal fluctuations in the disassembly of highly
excited nuclei are observed
\cite{Campi,Chomaz_1,Natowitz_1,DAgostino_1}. Fisher Droplet Model
analyses have been applied to extract critical parameters which
are very close to those observed for liquid-gas phase transitions
in macroscopic systems \cite{Fisher}. In previous studies on
liquid gas phase change, attention focused mostly on the larger
atomic nuclei. However, the larger Coulomb effects make it
difficult to reach the critical point of the finite nucleonic
matter. In contrary, it was suggested that the decreasing
importance of Coulomb effects makes the lightest nuclei the most
favorable venue for investigation of the critical point
\cite{Natowitz_2}.

In this paper,  we report results of an extensive investigation of
nuclear disassembly in nuclei of $A$ $\sim$ 36 excited  to
energies as high as 9 MeV/u and many observables are related to
the critical behavior at an excitation energy of 5.6$\pm$0.5 MeV.

\section{RESULTS AND DISCUSSIONS}

Using the TAMU NIMROD detector  and beams from the TAMU K500
Super-conducting Cyclotron, we have probed the properties of
excited quasi-projectile  (QP) fragments produced in the reactions
of 47 MeV/u $^{40}$Ar + $^{58}$Ni. Earlier work on systems at
energies near the Fermi energy have demonstrated the essential
binary nature of such collisions, even at relatively small impact
parameters. As a result, these collisions prove to be very useful
in preparing highly excited light nuclei.

The charged particle detector array of NIMROD which is set inside
a neutron ball includes 166 individual CsI detectors arranged in
12 rings in polar angles from $\sim$ $3^\circ$  to $\sim$
$170^\circ$. In these experiments each forward ring included two
super-telescopes composed of two Si-Si-CsI detectors and three
Si-CsI telescopes to identify intermediate mass fragments. In the
CsI detector H and He isotopes are clearly identified and Li
fragments are also isolated from the heavier fragments. In the
super-telescopes, all isotopes with atomic number $Z\leq 8$ are
clearly identified and in all telescopes particles are identified
by atomic number. The NIMROD neutron ball, which surrounds the
charged particle array, was used to determine the neutron
multiplicities for selected events. The correlation of the charged
particle multiplicity and the neutron multiplicity was used to
select violent collisions.

A new method to reconstruct QP has been developed in this work. We
first obtain the laboratory energy spectra for different LCP at
different laboratory angles and reproduce them using the three
source fits, {\it i.e.} the QP, NN and QT sources. From these fits
we know the relative contributions from the QP, NN and QT sources.
Employing this information to determine the energy and angular
dependent probabilities we analyze the experimental events once
again and, on an event by event basis, use a Monte Carlo sampling
method to assign each LCP, {\it i.e.}, to one of the sources QP,
or NN, or QT. For intermediate mass fragments (IMF) with Z$\geq$4,
we have used a rapidity cut ($>0.65$ beam rapidity) to assign IMF
to the QP source.  Once we have identified all LCPs and IMFs which
are assumed to come from the QP source, we can reconstruct the
whole QP source on an {\it event-by-event} basis. For the present
analysis we have selected reconstructed QP events with total
charge number $Z_{QP}$ $\geq$ 12 from violent collisions. The QP
source velocity was determined from momentum conservation of all
QP detected particles. The excitation energy distribution was
deduced using the energy balance equation.

\begin{figure}[htb]
\begin{minipage}[t]{80mm}
\includegraphics[scale=0.4]{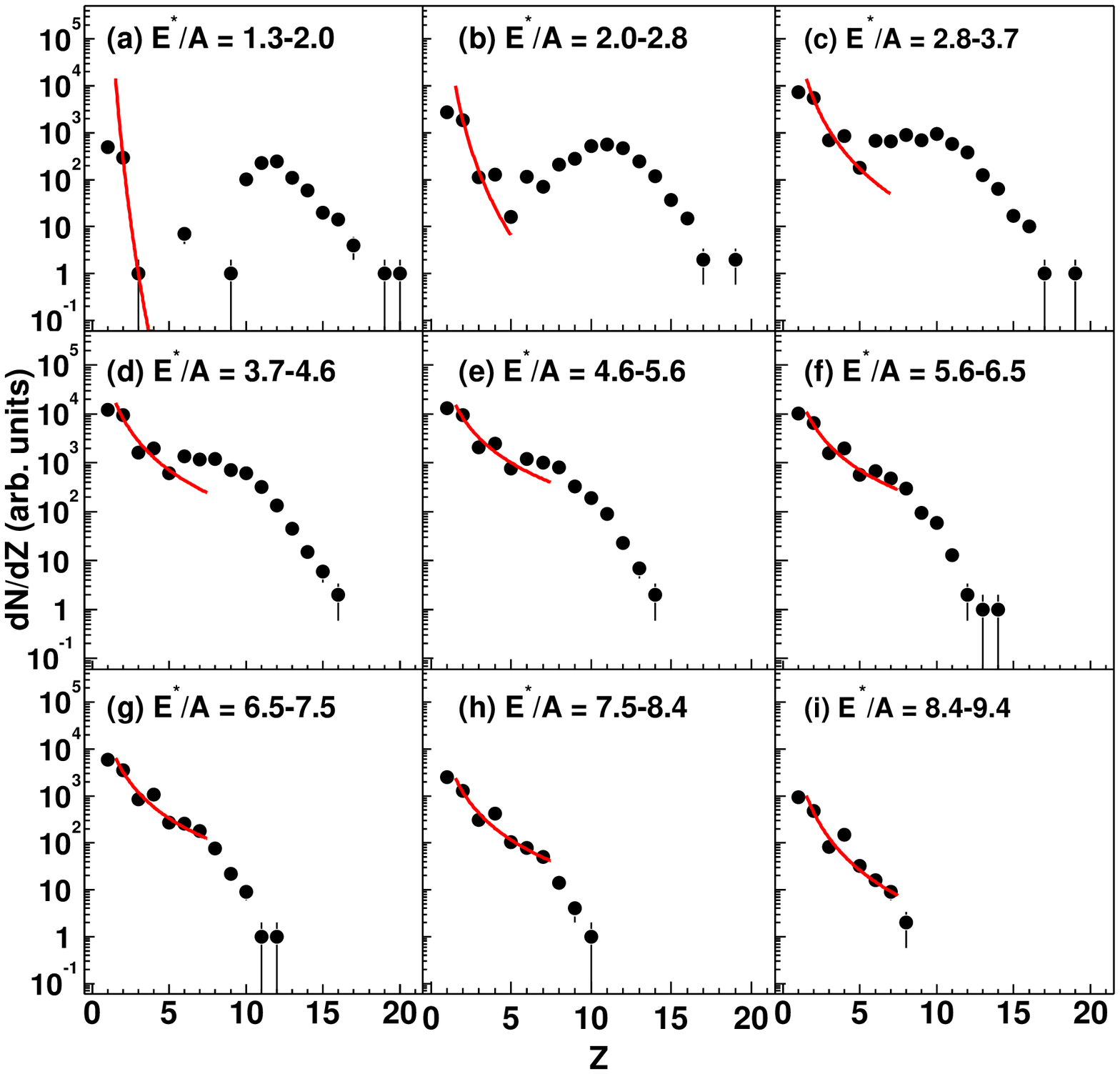}
\vspace{-1.5cm} \caption{Charge distribution of QP in different
$E^*/A$ window for the reaction  $^{40}$Ar + $^{58}$Ni. Lines
represent fits. }
 \vspace{-2.5cm} \label{fig_Zdist}
\end{minipage} 
\hspace{\fill}
\begin{minipage}[t]{80mm}
\includegraphics[scale=0.4]{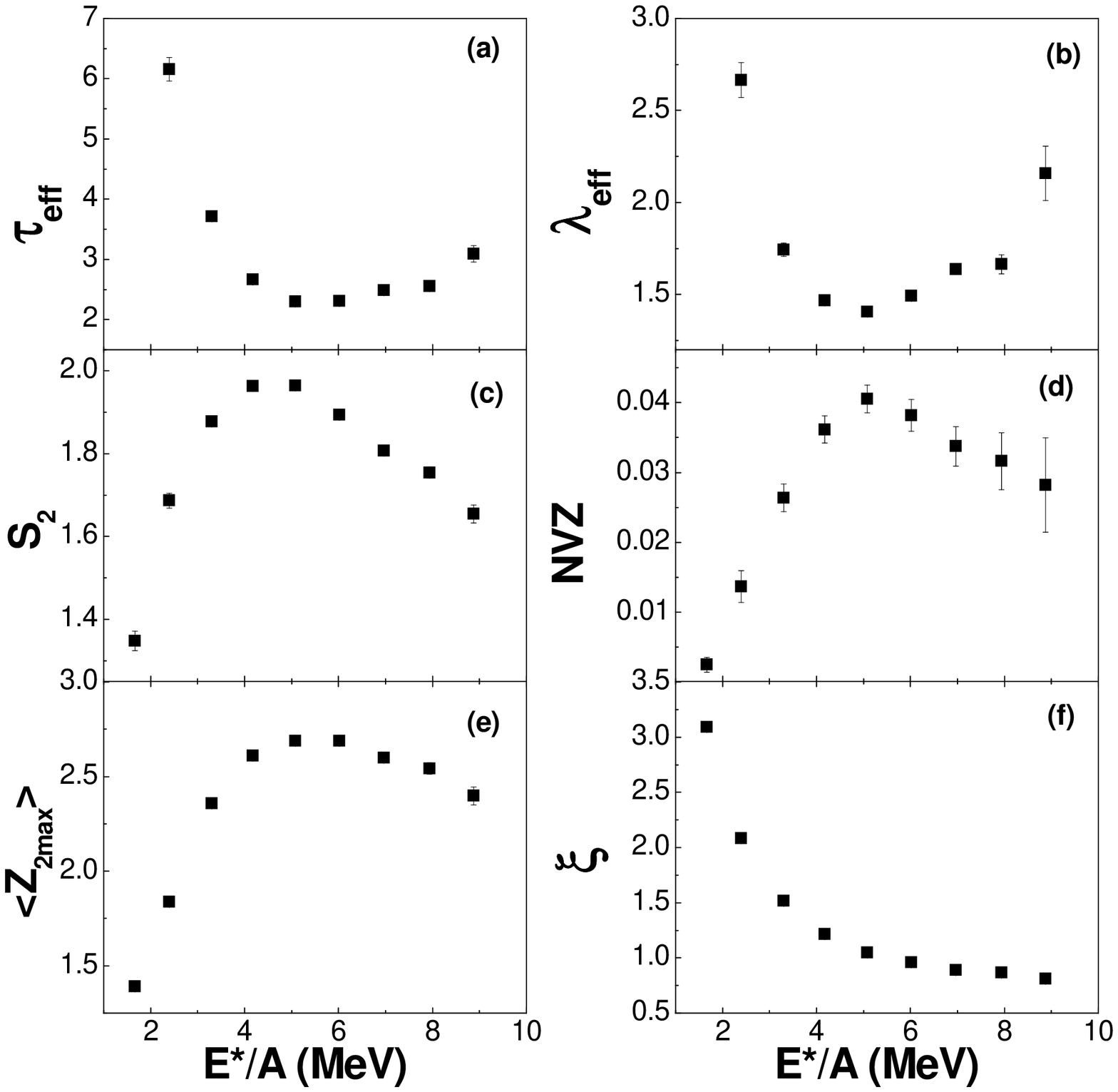}
\vspace{-2.5cm}\caption{The effective Fisher-law parameter
($\tau_{eff}$) (a), the effective exponential law parameter
($\lambda_{eff}$) (b), $\langle S_2\rangle$ (c), NVZ fluctuation
(d),   the mean charge number of the second largest fragment
$\langle Z_{2max}\rangle$ (e), the Zipf-law parameter $\xi$ (f).
See details in text.}
 \label{fluctuation}
\end{minipage}
\end{figure}

The Fisher droplet model has been extensively applied to the
analysis of multifragmentation. It predicts that relative yields
of fragments with $3 \leq Z \leq 14$ could be well described by a
power law dependence $A^{-\tau}$ when the liquid gas phase
transition occurs and $\tau$ value is the minimum in that point.
In Fig.~\ref{fig_Zdist} we present,  for the QP from the reactions
of $^{40}Ar$ + $^{58}Ni$,  yield distributions,  $dN/dZ$, observed
for our nine intervals of excitation energy. At low  excitation
energy a large $Z$ residue always remains, $i.e.$ the nucleus is
basically in the liquid phase accompanied by some evaporated light
particles. When $E^*/A$ reaches  $\sim$ 6.0 MeV/u, this residue is
much less prominent. As $E^*/A$ continues to increase, the charge
distributions become steeper, which indicates that the system
tends to vaporize. To quantitatively pin down the possible phase
transition point, we use a power law fit to the QP charge
distribution in the range of $Z$ = 2 - 7  to extract the effective
Fisher-law parameter $\tau_{eff}$ by $dN/dZ \sim Z^{-\tau_{eff}}$.
The Fig. 2(a) shows $\tau_{eff}$ vs excitation energy, a minimum
with $\tau_{eff}$ $\sim$ 2.3 is seen to occur in the $E^*/A$ range
of  5 to 6 MeV/u. $\tau_{eff}$ $\sim$ 2.3, is close to the
critical exponent of the liquid gas phase transition universality
class as predicted by Fisher's Droplet model \cite{Fisher}. The
observed minimum  is rather broad. While, assuming that the
heaviest cluster in each event represents the liquid phase, we
have attempted to isolate the gas phase by event-by-event removal
of the heaviest cluster from the charge distributions. We find
that the resultant distributions are better described as
exponential form $ exp(-\lambda_{eff} Z)$. The fitting parameter
$\lambda_{eff}$  was derived and is plotted against excitation
energy in Fig.~\ref{fluctuation}(b). A minimum is seen in the same
region where $\tau_{eff}$ shows a minimum.

\vspace{-2.7cm}

To further explore this region we have investigated other proposed
observables commonly related to fluctuations and critical
behavior. Fig. 2(c) shows the mean normalized second moment
\cite{Campi}, $\langle S_2\rangle$ as a function of excitation
energy. A peak is seen around 5.6 MeV/u, it indicates that the
fluctuation of the fragment distribution is the largest in this
excitation energy region. Similarly,  the normalized variance in
$Z_{max}/Z_{QP}$ distribution (i.e. NVZ =
$\frac{\sigma^2_{Z_{max}/Z_{QP}}}{\langle Z_{max}/Z_{QP}\rangle}$)
\cite{Dorso} shows a maximum in the same excitation energy region
[Fig. 2(d)], which illustrates the maximal fluctuation for the
largest fragment (order parameter) is reached around $E^*/A$ = 5.6
MeV. Except the largest fragment, the second largest fragment also
shows its importance in the above turning point. Fig. 2(e) shows a
broad peak of $\langle Z_{2max}\rangle$ - the average atomic
number of the second largest fragment exists at 5.6 MeV/u.

The significance of the 5-6 MeV region in our data is further
indicated by a Zipf's law analysis proposed by Ma \cite{Ma_1}. In
such an analysis, the cluster size is employed as the variable to
make a Zipf-type plot. We can define a  Zipf-type plot by plotting
the mean sizes of fragments which are rank-ordered in size, i.e.,
largest, second largest, etc. as a function of  their rank
\cite{Ma_1}. The resultant distributions are fitted with a power
law, $\langle Z_{rank}\rangle \propto  rank^{-\xi}$, where $\xi$
is the Zipf's law parameter. When $\xi \sim 1$, Zipf's law is
satisfied.  In this case, the mean charge of the second largest
fragment is 1/2 of that of the the largest fragment;  that of the
third largest fragment is 1/3 of the largest fragment, etc. This
is a kind of special topological fragment structure. By the fits
we obtained parameters $\xi$ as a function of excitation energy as
shown in Fig.2(e). This rank ordering of the probability
observation of fragments of a given atomic number, from  the
largest to the smallest, does indeed lead to a Zipf power law
parameter $\xi$ $\sim$ 1 in  the 5-6 MeV/u range.

The calculations of the phase transition models,  namely the
lattice gas model (LGM) and statistical multifragmentation model
(SMM), show very similar behavior as the experimental data in
Fig.1 and 2. But the sequential decay model fails. This indicates
that the data around $E^*/A$ = 5.6 MeV/u can be actually explained
by the phase change \cite{Ma_2}.

\section{SUMMARY}
In our measurements for the disassembly  of small nucleui with A
$\sim 36$, the  maximal fluctuations are observed at $5.6\pm0.5$
MeV/u excitation energy. Also the fragment topological structures
suggest the onset of a phase change. Comparisons with results of
LGM and SMM  calculations suggest that the critical point may have
been reached. Taken together, this body of evidence suggests a
phase change at, or extremely close to, the critical point of this
light nuclear system.

{\bf Acknowledgements:} The work was supported by the the U.S.
Department of Energy and the Robert A. Welch Foundation under
Grant No. A330. The work of YGM was partially supported by the
Major State Basic Research Development Program in China under
Contract No. G2000077404.

\end{document}